\documentclass[twoside,reqno]{HERON}
\usepackage{epsfig,cite,colordvi}
\usepackage{graphicx}
\usepackage{url}
\usepackage{amssymb,amsmath,amscd,epsf}
\usepackage{times}
\usepackage{makeidx}
\def\be{\begin{equation}}
\def\lan{\left\langle}
\def\ran{\right\rangle}
\def\ee{\end{equation}}
\def\barr{\begin{array}}
\def\earr{\end{array}}

\def\l{\left}
\def\r{\right}
\def\dis{\displaystyle}
\def\ed{\end{document}}

\def\co{{\cal O}}
\def\ch{{\cal H}}
\def\can{{\cal N}}

\def\cg{{\cal G}}
\def\ce{{\cal E}}

\def\cn{{\bf n}}

\def\tmp{\widetilde{m_p}}
\def\tmn{\widetilde{m_n}}

\def\wm{{\widetilde {m}}}

\begin{document}

\title{Spectral distribution Method for neutrinoless double  beta decay:
Results for $^{82}$Se and $^{76}$Ge}

\runningheads{Spectral distribution Method for neutrinoless double beta
decay}{V.K.B. Kota, R.U. Haq}

\begin{start}

\author{V.K.B. Kota}{1}, \coauthor{R.U. Haq}{2}

\index{Kota, V.K.B.}
\index{Haq, R.U.}

\address{Physical Research Laboratory, Ahmedabad 380009, India}{1}

\address{Department of Physics, Laurentian Unversity, Sudbury, ON P3E 2C6,
Canada}{2}

\begin{Abstract}

Statistical spectral distribution method based on shell model and random matrix
theory is developed for calculating neutrinoless double beta decay nuclear
transition matrix elements. First results obtained for $^{82}$Se and $^{76}$Ge
using the spectral method are close to the available shell model results. 

\end{Abstract}
\end{start}

\section{Introduction}

Neutrinoless double beta decay ($0\nu \beta \beta$ or NDBD) which involves
emission of two electrons without the accompanying neutrinos and which violates 
lepton number conservation has been an important and challenging problem both 
for the experimentalists and theoreticians.  Recent neutrino oscillation 
experiments have demonstrated that neutrinos have mass
\cite{Kamok,SNO-1,SNO-2}.  The observation of $0\nu \beta \beta$ decay is
expected to provide information regarding  the absolute  neutrino mass which is,
as yet, not known. As a result, experimental programs to observe this decay
have  been initiated at different laboratories across the globe and  already are
in advanced stages of development. The most recent results for $0\nu \beta
\beta$ decay of $^{136}$Xe have been reported by EXO-200  collaboration
\cite{Exo} and KamLand-Zen  collaboration \cite{Kam}. They give a lower limit of
$3.4\times 10^{25}$ yr for the half-life.  Further, phase I results from GERDA
experiment \cite{Gerda} for $^{76}$Ge  gives a lower limit of $3.0\times
10^{25}$ yr for the half-life.  Nuclear transition matrix elements (NTME) are
the essential ingredient for extracting the neutrino mass from the half lives
\cite{Vogel}. There has been considerable effort to obtain NTME for various
candidate nuclei and they have been calculated  theoretically using a variety of
nuclear models: (i) large scale shell model; (ii) quasi-particle random phase
approximation and its variants;  (iii) proton-neutron interacting boson model;
(iv) particle number and angular momentum projection including configuration
mixing within the generating coordinate method framework; (v) projected
Hartree-Fock-Bogoliubov  method with pairing plus  quadrupole-quadrupole
interaction. A detailed comparative study of  the results from these various
methods is discussed in \cite{IBM2,JS1}. In addition, more recently the  so
called deformed shell model based on Hartree-Fock single particle states has
been used for the candidate nuclei in the A=60-90 region \cite{KS}. It is
important to note that the predictions of various models for NTME vary typically
from  2 to 6 \cite{IBM2}.

The statistical spectral distribution method (SDM) developed by French and
collaborators for nuclear structure is well documented \cite{KH}  and the
operation of embedded Gaussian orthogonal ensemble of random matrices (EGOE) in
nuclear shell model spaces forms the basis for SDM \cite{Ko-b}.  With this, it
is natural that one should develop and apply SDM for calculating the NTME for
$0\nu \beta \beta$ and compare the results with those obtained using shell model
and other models. This is addressed in the present paper with first results for
$^{82}$Se and $^{76}$Ge. The essential point is that NTME can be viewed as a
transition  strength (square of the matrix element connecting a given initial
state to a final state by a transition operator) generated by the NDBD operator
that is two-body in nature. Therefore, SDM for transition strengths as given in
\cite{FKPT,KM,Ko-15,KH} can be used as the starting point for further
developments and applications. Let us add that SDM is sometimes called moment
method. 

We now give a preview.   Section 2 gives a brief discussion of the relation
between neutrino mass and NTME and then deals with the structure of the NDBD
transition operator.  Section 3 deals with the  details of SDM for transition
strengths as applicable for NDBD. In Section 4, we present SDM results for
$^{82}$Se and $^{76}$Ge NDBD NTME. For these two nuclei, experiments SuperNEMO
and GERDA+MAJORANA respectively are under development to measure NDBD half
lives. Finally,  conclusions and future outlook are given in Section 5. 

\section{Neutrinoless double beta decay and NTME}

In $0\nu \beta \beta$, the half-life for the 0$^+_i$ ground state (gs) of a
initial even-even nucleus decay to the 0$^+_f$ gs of the final even-even nucleus
is  given by \cite{Vogel}
\be
\l[ T_{1/2}^{0\nu}(0^+_i \to 0^+_f) \r]^{-1} =  G^{0\nu}
\l| M^{0\nu} (0^+)\r|^2 \l(\dis\frac{\lan m_\nu \ran}{m_e}\r)^2 \;,
\label{eq.dbd1}
\ee
where $\lan m_\nu \ran$ is the effective neutrino mass (a combination of
neutrino mass eigenvalues and also involving the neutrino mixing matrix).  The
$G^{0\nu}$ is a phase space integral  (kinematical factor); tabulations for 
$G^{0\nu}$ are available. The $M^{0\nu}$ represents NTME of the NDBD  transition
operator and it is a sum of a Gamow-Teller like ($M_{GT}$), Fermi like ($M_F$)
and tensor ($M_T$) two-body operators. Since it is well known that the  tensor
part contributes only up to 10\% of the matrix elements,  we will neglect the
tensor part. Then, from the closure approximation which is well justified for
NDBD, we have
\be
\barr{rcl}
M^{0\nu} (0^+) & = & M^{0\nu}_{GT} (0^+) - \dis\frac{g_V^2}{g_A^2} 
M^{0\nu}_{F} (0^+) = \lan 0^+_f \mid\mid \co(2:0\nu) \mid\mid 0^+_i \ran \;,
\\
\co(2:0\nu) & = & \dis\sum_{a,b} \ch(r_{ab}, \overline{E}) 
\tau_a^+ \tau_b^+ \l( \sigma_a \cdot \sigma_b- \dis\frac{g_V^2}{g_A^2} \r)
\;.
\earr \label{eq.dbd2}
\ee
As seen from Eq. (\ref{eq.dbd2}),  NDBD half-lives are generated by the two-body
transition operator $\co(2:0\nu)$; note that $a,b$ label  nucleons. The $g_A$
and $g_V$ are the weak axial-vector and vector coupling constants.  The
$\ch(r_{ab}, \overline{E})$ in Eq. (\ref{eq.dbd2}) is called the `neutrino
potential'. Here, $\overline{E}$ is the average energy of the virtual
intermediate states used in the closure approximation. The form given by Eq.
(\ref{eq.dbd2}) is justified {\it only if the exchange of the light majorana
neutrino is indeed the mechanism responsible for the NDBD}. With the phase space
factors fairly well known, all one needs are NTME $|M^{0\nu} (0^+)|=\l|\lan
0^+_f \mid\mid \co(2:0\nu) \mid\mid 0^+_i \ran\r|$. Then, measuring the 
half-lives makes  it possible to deduce neutrino mass using Eq. (\ref{eq.dbd1}).

The neutrino potential is of the form $\ch(r_{ab}, \overline{E}) =
[R/r_{ab}]\,\Phi(r_{ab},\overline{E})$ where $R$ in fm units is the nuclear
radius and similarly  $r_{ab}$ is in fm units. A simpler form for the function
$\Phi$, involving sine and cosine integrals, as given in \cite{Vogel} and
employed in \cite{KS}, is used in the present work. It is useful to note that
$\Phi(r_{ab},\overline{E}) \sim  \exp ({-\frac{3}{2} \frac{\overline{E}}{\hbar
c} r_{ab}})$.   The effects of short-range correlations in the wavefunctions are
usually taken into  account by multiplying the wavefunction by the Jastrow
function $[1 - \gamma_3e^{-\gamma_1 r_{ab}^2}  ( 1 - \gamma_2 r_{ab}^2 )]$.
There are other approaches \cite{Shoun} for taking into account the short range
correlations  but they are not considered here. Now, keeping the wavefunctions
unaltered, the Jastrow  function can be incorporated into $\ch(r_{ab},
\overline{E})$ giving an effective $\ch_{eff}(r_{ab}, \overline{E})$,
\be
\ch(r_{ab}, \overline{E}) \to 
\ch_{eff}(r_{ab}, \overline{E}) = \ch(r_{ab}, \overline{E}) 
[ 1 - \gamma_3 \;e^{-\gamma_1 \; r_{ab}^2} ( 1 - \gamma_2 \; r_{ab}^2 ) 
]^2 \;.
\label{eq.dbd3b}
\ee 
The choice of the values for the parameters $\gamma_1$, $\gamma_2$ and 
$\gamma_3$ is given in Section 4.

Let us say that for the nuclei under consideration, protons are in the single
particle (sp) orbits $j^p$ and similarly neutrons in $j^n$. Using the usual
assumption that the radial part of the sp states are those of the harmonic
oscillator, the proton sp states are completely specified by
($\cn^p,\ell^p,j^p$) with $\cn^p$ denoting oscillator radial quantum number so
that for a oscillator shell $\can^p$, $2\cn^p+\ell^p=\can^p$. Similarly, the
neutron sp states are ($\cn^n,\ell^n,j^n$). In terms of the creation
($a^\dagger$) and annihilation ($a$) operators, normalized two-particle
(antisymmetrized) creation operator $A^J_\mu(j_1j_2) = 
(1+\delta_{j_1j_2})^{-1/2}  (a^\dagger_{j_1}a^\dagger_{j_2})^J_\mu$ and then
$A^J_\mu \l|0\ran = \l|(j_1 j_2)J \mu\ran$ represents a normalized  
two-particle state. At this stage, it is important to emphasize that we are
considering only $0^+$ to $0^+$ transitions in $0 \nu \beta \beta$ and therefore
only the $J$ scalar part of $\co(2:0\nu)$ will contribute to $M^{0\nu}$. With
this, the NDBD  transition operator  can be written as,
\be
\co(2:0\nu) = \dis\sum_{j_1^p \geq j_2^p;j_3^n \geq j_4^n;J} \co_{j_1^p \, 
j_2^p; j_3^n \, j_4^n}^J (0\nu) \dis\sum_\mu A^J_\mu(j_1^pj_2^p) \l\{
A^J_\mu(j_3^nj_4^n) \r\}^\dagger \;.
\label{eq.dbd4}
\ee  
Here, $\co_{j_1^p \, j_2^p; j_3^n \, j_4^n}^J (0\nu) = \lan (j_1^p \,  j_2^p) J
M \mid \co(2:0\nu)  \mid (j_3^n \, j_4^n) J M \ran_a$ are two-body matrix
elements (TBME) and \lq{$a$}\rq denotes antisymmetrized two-particle
wavefunctions; $J$ is even  for $j_1=j_2$  or $j_3=j_4$. The TBME are obtained
by using the standard approach based on Brody-Moshinisky  brackets and Talmi
integrals.  

\section{Spectral distribution method for NDBD}

\subsection{State densities and Gaussian form}

Let us consider shell model sp orbits $j^p_1,  j^p_2, \ldots, j^p_r$ with $m_p$
protons distributed in them. Similarly, $m_n$ neutrons are distributed in
$j^n_1, j^n_2, \ldots, j^n_s$ orbits.  Then, the proton configurations are
$\tmp=[m_p^1, m_p^2, \ldots, m_p^r]$ where $m_p^i$ is number of protons in the
orbit $j_i^p$ with $\sum_{i=1}^r\,m_p^i=m_p$. Similarly, the neutron
configurations are $\tmn=[m_n^1, m_n^2, \ldots, m_n^s]$ where $m_n^i$ is number
of neutrons in the orbit $j_i^n$ with $\sum_{i=1}^s\,m_n^i=m_n$. With these,
$(\tmp, \tmn)$'s denote proton-neutron configurations.  The nuclear effective
Hamiltonian is one plus two-body, $H=h(1)+V(2)$ and we  assume that the one-body
part $h(1)$ includes the mean-field producing part of  the two-body interaction.
Thus, $V(2)$ is the irreducible two-body part of $H$ \cite{KH}. From now on, for
simplicity we shall denote $h=h(1)$ and $V=V(2)$. The state density $I^H(E)$, 
with  $\lan \lan -- \ran \ran$ denoting trace, can be written as a sum of the
partial densities defined over $(\tmp, \tmn)$, 
\be
\barr{l}
I^{(m_p, m_n)}(E) = \lan \lan \delta(H-E) \ran \ran^{(m_p, m_n)} =
\dis\sum_{(\tmp, \tmn)}\, \lan \lan \delta(H-E) \ran \ran^{(\tmp, \tmn)} \\
\\
= \dis\sum_{(\tmp, \tmn)}\,I^{(\tmp, \tmn)}(E) = \dis\sum_{(\tmp, \tmn)}\,
d(\tmp,\tmn)\,
\rho^{(\tmp, \tmn)}(E)\;.
\earr \label{eq.ndbd1}
\ee
Here, $d(\tmp,\tmn)$ is the dimension of the configuration $(\tmp, \tmn)$ and 
$\rho^{(\tmp, \tmn)}(E)$ is normalized to unity. For strong enough two-body
interactions (this is valid for nuclear interactions \cite{KH}), the operation 
of embedded GOE of one plus two-body interactions [EGOE(1+2)] will lead to 
Gaussian form for the partial densities $\rho^{(\tmp, \tmn)}(E)$ and therefore,
\be
I^{(m_p, m_n)}(E) = \dis\sum_{(\tmp, \tmn)}\;I^{(\tmp, \tmn)}_\cg(E)\;.
\label{eq.ndbd2}
\ee
In Eq. (\ref{eq.ndbd2}), $\cg$ denotes Gaussian. The Gaussian partial densities
are defined by the centroids $E_c(\tmp, \tmn)=\lan H \ran^{(\tmp, \tmn)}$ and
variances $\sigma^2(\tmp, \tmn) = \lan H^2  \ran^{(\tmp, \tmn)} -
[E_c(\tmp,\tmn)]^2$. Expressions for these follow easily from trace propagation
methods \cite{CFT,KH}. In practical applications to nuclei, Eq. (\ref{eq.ndbd2})
has to be applied in fixed-$J$ spaces \cite{Zel-1,Zel-2} or an approximate $J$
projection has to be carried out \cite{KH,Ko-96,Fr-06}. We will return to this
question in Section 3.4.

\subsection{Transition strength densities and bivariate Gaussian form}

Given a transition operator $\co$, the spectral distribution method for 
transition strengths starts with the transition strength density 
$I^H_\co(E_i, E_f)$,
\be
I^H_\co(E_i, E_f) = I(E_f) |\lan E_f \mid \co \mid E_i \ran|^2 I(E_i)
\label{eq.ndbd3}
\ee
where $E$'s are eigenvalues of $H$. A plausible way to proceed now \cite{FKPT}
is to first construct the transition strength density with $H=h=\sum_r
\epsilon_r n_r$; $n_r$ is the number operator  for the orbit $r$ and $\epsilon_r$
are the sp energies (spe).  As the configurations $(\tmp, \tmn)$ are eigenstates
of $h$, it is straightforward to construct $I^{h}_{\co}$ \cite{KH}.
Next the interaction $V$, the two-body part of $H$, is switched on. Then, the
role of $V$ is to  locally spread $I^{h}_{\co}$ and therefore the strength
density will be a  bivariate convolution of $I^{h}_\co$ and $\rho^{V}_{\co}$;
the spreading function (normalized to unity)  $\rho^{V}_{\co}$ is a 
bivariate distribution.  For strong enough interactions, operation of EGOE(1+2)
generates bivariate Gaussian form for $\rho^{V}_{\co}$ and this result has been
established for NDBD type operators in \cite{Ko-15}. Applying this, with some
additional approximations as discussed ahead, will give
\be
\barr{l}
\l|\lan E_f \mid \co \mid E_i \ran\r|^2 = \dis\sum_{\wm_i,\wm_f} 
\dis\frac{
I^{\wm_i}_{\cg}(E_i) I^{\wm_f}_{\cg}(E_f)}{I^{m_i}(E_i) I^{m_f}(E_f)}
\l|\lan \wm_f \mid \co \mid \wm_i\ran\r|^2 \\ 
\times \dis\frac{\rho^V_{\co : biv-\cg}(E_i , E_f , 
\ce_{\co:V}(\wm_i), \ce_{\co:V}(\wm_f),
\sigma_{\co:V}(\wm_i), \sigma_{\co:V}(\wm_f), \zeta_{\co:V}(\wm_i , \wm_f))}{
\rho_{\cg}^{\wm_i}(E_i)
\rho_{\cg}^{\wm_f}(E_f)}\;; \\ 
\l|\lan \wm_f \mid \co \mid \wm_i\ran\r|^2 = \l[d(\wm_i) d(\wm_f) \r]^{-1}\;
\dis\sum_{\alpha , \beta} 
\l|\lan \wm_f, \alpha \mid \co \mid \wm_i, \beta \ran\r|^2\;.
\earr \label{eq.strn-app3}
\ee
This is the basic equation that allows one to use SDM for the calculation of
NTME $M^{0\nu}$. In order to apply this, we need the marginal centroids
$\ce_{\co:V}(\wm_i)$ and $\ce_{\co:V}(\wm_f)$, marginal variances
$\sigma^2_{\co:V}(\wm_i)$ and $\sigma^2_{\co:V}(\wm_f)$ and the correlation
coefficient $\zeta_{\co:V}(\wm_i , \wm_f)$ defining $\rho^{V}_{\co :biv-\cg}$. 
Also, we need  $\l|\lan \wm_f \mid \co \mid \wm_i\ran\r|^2$. Note that $\wm = 
(\tmp, \tmn)$ in actual applications and further, the angular momentum quantum
numbers for the parent and daughter nuclei involved in $0\nu \beta \beta$ decay
need to be considered. We will turn to these now.

\subsection{SDM for NTME for $0\nu \beta \beta$ decay}

Firstly, the marginal centroids and variances in Eq. (\ref{eq.strn-app3}) are
approximated, following random matrix theory \cite{Ko-15}, to the  corresponding
state density centroids and variances  (see for example \cite{KM,FKPT}) giving
$\ce_{\co:V}((\tmp, \tmn)_r) \approx E_c((\tmp, \tmn)_r) =  \lan H \ran^{(\tmp,
\tmn)_r}$ and $\sigma^2_{\co:V}((\tmp, \tmn)_r)) \approx  \sigma^2((\tmp,
\tmn)_r)) = \lan V^2 \ran^{(\tmp, \tmn)_r}$; $r=i,f$.  For the correlation
coefficient $\zeta$, there is not yet any valid form involving configurations.
Therefore, the only plausible way forward currently is to estimate $\zeta$ as a
function of $(m_p,m_n)$ using random matrix theory given in \cite{Ko-15,FKPT}.
Then, the definition of $\zeta$ is
\be
\barr{l}
\zeta_{\co:V}(m_p,m_n) = \\
\dis\frac{\lan \co(2: 0\nu)^{\dagger}\, V \, 
\co(2: 0\nu)\, 
V \ran^{(m_p,m_n)}}{\dis\sqrt{\lan \co(2: 0\nu)^{\dagger}\,V^2\,
\co(2: 0\nu)\ran^{
(m_p,m_n)}\;\lan \co(2: 0\nu)^{\dagger}\,\co(2: 0\nu)\,
V^2\ran^{(m_p,m_n)}}} \;.
\earr \label{eq.strn-app5}
\ee
Results in Sections 5 and 6 of \cite{Ko-15}, obtained using EGOE representation
for both the $\co$ and $V$ operators, will allow one to obtain $\zeta(m_p,m_n)$.
For nuclei of interest, using the numerical results in Table 2 of \cite{Ko-15},
it is seen that $\zeta \sim 0.6-0.8$. These values are used in the $M^{0\nu}$
calculations reported in Section 4 ahead.

In order to apply Eq. (\ref{eq.strn-app3}), in addition to the marginal
centroids, variances and $\zeta$, we also need an expression for $\l|\lan \wm_f
\mid \co \mid \wm_i\ran\r|^2$, the configuration mean square matrix
element of the transition operator.  Applying the propagation theory given in 
\cite{CFT} will give,
\be
\barr{l}
\l|\lan (\tmp, \tmn)_f \mid \co(2: 0\nu) \mid (\tmp, 
\tmn)_i \ran\r|^2 = \l\{ d[(\tmp, \tmn)_f]\r\}^{-1} 
\\ 
\times \;\dis\sum_{\alpha , \beta , \gamma ,\delta}\;
\dis\frac{m^i_n(\alpha)\, [m^i_n(\beta)-\delta_{\alpha \beta}]\, 
[N_p(\gamma) - m^i_p(\gamma)]\, [N_p(\delta) -
m^i_p(\delta) - \delta_{\gamma \delta}]}{N_n(\alpha)\, [N_n(\beta) -
\delta_{\alpha \beta}]\, N_p(\gamma)\, [N_p(\delta) - \delta_{\gamma \delta}]}
\\ 
\times \dis\sum_{J_0} \; \l[\co^{J_0}_{\gamma^p \delta^p \alpha^n
\beta^n}(0\nu)\r]^2
(2J_0 +1) \;;\\ 
(\tmp, \tmn)_f = (\tmp, \tmn)_i \times \l(1^+_{\gamma_p} 1^+_{\delta_p}
1_{\alpha_n} 1_{\beta_n} \r) \;.
\earr \label{eq.strn-app6}
\ee
Note that in Eq. (\ref{eq.strn-app6}), the final configuration is defined by
removing  one neutron from orbit $\alpha$ and another from $\beta$ and then
adding one proton in orbit $\gamma$ and another in orbit $\delta$. Also,
$N_p(\alpha)$ is the degeneracy of the proton orbit $\alpha$ and similarly
$N_n(\gamma)$ for the neutron orbit $\gamma$.   

\subsection{Angular momentum decomposition of transition strengths}

For NDBD NTME calculations, to complete the transition strength theory given by
Eqs. (\ref{eq.strn-app3}) -  (\ref{eq.strn-app6}), we need $J$ projection as the
quantity of interest is 
$$
\l|\lan E_f J_f=0\mid \co \mid E_i J_i=0\ran\r|^2
$$
where $E_i$ and $E_f$ are the ground state energies of the parent and daughter
nuclei respectively and similarly $J_i$ and $J_f$. Firstly, note that
\be
\barr{l}
\l|\lan E_f J_f=0\mid \co(2: 0\nu) \mid E_i J_i=0\ran\r|^2 \\
= \dis\frac{\lan\lan [\co(2: 0\nu)]^\dagger X(H,J^2,E_f,J_f) \co(2: 0\nu) 
Y(H,J^2,E_i,J_i) \ran\ran^{(m_p^i m_n^i)}}{\lan\lan X(H,J^2,E_f,J_f)
\ran\ran^{(m_p^f m_n^f)}  \lan\lan Y(H,J^2,E_i,J_i) \ran\ran^{(m_p^i m_n^i)}} 
\\
=\dis\frac{I^{(m_p^i m_n^i),(m_p^f m_n^f)}_{\co(2: 0\nu)}(E_i,E_f,J_i,J_f)}{
I^{(m_p^i m_n^i)}(E_i,J_i) I^{(m_p^f m_n^f)}(E_f,J_f)} \;; \\
X(H,J^2,E_f,J_f) =  \delta(H-E_f) \delta(J^2 -J_f(J_f+1))\;,\\
Y(H,J^2,E_i,J_i) = \delta(H-E_i) \delta(J^2-J_i(J_i+1)) \;.
\earr \label{eq.strn-app7}
\ee
The four variate density $I_{\co}(E_i,E_f,J_i,J_f)=I_{\co}(E_i,E_f)
\rho_{\co}(J_i,J_f : E_i,E_f)$ where $\rho$ is a conditional density. Now, using
the fact that $J_f$ ($J_i$) is uniquely determined by $J_i$ ($J_f$) for the
$\co(2: 0\nu)$ operator and the $J$-factoring used in \cite{FKPT} will give the
approximation
\be 
I_{\co(2: 0\nu)}(E_i,E_f,J_i,J_f) \sim I_{\co(2: 0\nu)}(E_i,E_f)
\sqrt{C_{J_i}(E_i) C_{J_f}(E_f)}\;. 
\label{eq.strn-dum1} 
\ee 
The function $C_J(E)$ involves spin cut-off factor as given below. In addition,
we have the well established result \cite{CFT,Fr-06,FKPT,KH} $I(E,J) =
I(E) C_{J}(E)$. Using these will give, 
\be
\barr{l} 
\l|\lan E_f J_f=0\mid \co(2: 0\nu) \mid E_i J_i=0\ran\r|^2  =
\dis\frac{\l|\lan E_f \mid \co(2:0\nu) \mid E_i \ran\r|^2}{\dis\sqrt{
C_{J_i=0}(E_i) C_{J_f=0}(E_f)}}\; \;; \\ 
\\
C_{J_r}(E_r)=\dis\frac{(2J_r+1)}{\dis\sqrt{8\pi}\; \sigma^3_J(E_r)}  \exp
-(2J_r+1)^2/8\sigma_J^2(E_r) \stackrel{J_r=0}{\longrightarrow} \dis\frac{1}
{\dis\sqrt{8\pi} \sigma^3_J(E_r)} 
\earr \label{eq.strn-app8} 
\ee
where $r=i,f$. Note that $\sigma_J^2(E) = \lan J_Z^2\ran^E$ is the energy
dependent spin cut-off factor. In the approximation $C_{J_i=0}(E_i) \sim
[\sqrt{8\pi} \sigma^3_J(E_i)]^{-1}$ (similarly for $C_{J_f=0}(E_f)$), we have
used the fact that in general $\sigma_J(E) >> 1$. The spin cut-off factor can be
calculated using SDM \cite{KH,Ko-96,Fr-06}. Carrying this out for the nuclei of
interest in the present study, it is seen that $\sigma_J(E) \sim 3-4$ with $E$
varying up to  5 MeV excitation. Similarly, for lower $2p-1f$ shell nuclei 
studied in \cite{Ko-96},  $\sigma_J(E) \sim 4-6$. Because of the uncertainties
in using spin decomposition via Eq. (\ref{eq.strn-app8}), in the present work
$M^{0\nu}$ is calculated by varying $\sigma_J$ from 3 to 6. In principle it is
possible to avoid the use of spin cut-off factors (see Section 5).

\section{SDM results for $^{82}$Se and $^{76}$Ge $0\nu \beta \beta$ NTME} 

In the first application of SDM given in Section 3, we have chosen $^{82}$Se as
large shell model results, obtained using an easily available and well
established effective interaction, for the NTME for the $0\nu \beta\beta$ decay
to $^{82}$Kr are available in \cite{Se82}. In addition, SuperNEMO experiment
will be measuring $^{82}$Se $0\nu \beta\beta$ decay half-life \cite{dbd-expt1}.
In the shell model calculations, $^{56}$Ni is the core and the valence protons
and neutrons in $^{82}$Se and $^{82}$Kr occupy the $f_{5/2}pg_{9/2}$ orbits 
$^1p_{3/2}$, $^0f_{5/2}$, $^1p_{1/2}$ and $^0g_{9/2}$. The effective 
interaction used is JUN45. The spe and TBME defining JUN45 are given in
\cite{jun45}. In the SDM application, same shell model space, spe and TBME are
employed. Firstly, all the proton-neutron configurations are generated for both
$^{82}$Se and $^{82}$Kr. Number of positive parity configurations is 316 for
$^{82}$Se and 1354 for $^{82}$Kr. 

Using the formula in \cite{CFT} and the JUN45 interaction, the centroids and
variances defining the Gaussian partial densities in Eq. (\ref{eq.ndbd2}) are
calculated. These will also give the marginal centroids and variances in Eq.
(\ref{eq.strn-app3}). The average width ($\overline{\sigma}$) for $^{82}$Se
configurations is 3.34 MeV with a 9\% fluctuation. Similarly, for $^{82}$Kr,
$\overline{\sigma} = 4.7$ MeV with a 5\% fluctuation.  Proceeding further, the
TBME $\co_{j_1^p \,  j_2^p; j_3^n \, j_4^n}^J (0\nu)$ defining the $0\nu \beta
\beta$ transition operator are calculated and they are 259 in  number for the
chosen set of sp orbits. The choices made for the various parameters in the 
transition operator are (i) $R=1.2A^{1/3}$ fm; (ii) $b=1.003A^{1/6}$ fm; (iii)
$\overline{E}=1.12A^{1/2}$ MeV; (iv) $g_A/g_V=1$ (quenched); (v)
$\gamma_1=1.1\,fm^{-2}$, $\gamma_2=0.68\,fm^{-2}$ and $\gamma_3=1$ (these are
Miller-Spencer Jastrow correlation parameters). Then, applying Eq.
(\ref{eq.strn-app6}), the configuration mean square matrix elements of the
transition operator are obtained for all the configurations. With all these, in
order to apply Eqs. (\ref{eq.strn-app3}) and (\ref{eq.strn-app8}), we need the
ground states of $^{82}$Se and $^{82}$Kr and also the values of the $\zeta$ and
$\sigma_J$ parameters.

Using the so called Ratcliff procedure \cite{KH,KAR}, the ground states are
determined in SDM. For this one  needs a reference level with energy ($E_R$)
and angular momentum and parity $J^\pi$ value ($J_R^\pi$) and  also the total
number of states up to and including the reference level ($N_R$). The constraint
in choosing the reference level is that the $J^\pi$ values for all levels up to
the reference level should be known with certainty. Satisfying this, we have,
from the most  recent data \cite{nndc}, for $^{82}$Se the values $E_R=1.735$MeV,
$J_R^\pi =4^+$ and $N_R =21$. Similarly, for $^{82}$Kr we have $E_R=2.172$MeV,
$J_R^\pi =0^+$ and $N_R =34$. The ground states are found to be $\sim 3\sigma$
below the lowest configuration  centroid.

After obtaining the ground states, the ground to ground NTME are calculated 
using Eq. (\ref{eq.strn-app3}) with $J$-decomposition via Eq.
(\ref{eq.strn-app8}). For the correlation coefficient $\zeta$  the values $0.6$,
$0.65$, $0.7$ and $0.8$ are used as stated in Section 3.3. Similarly, assuming
$\sigma_J(E_i(gs)) = \sigma_J(E_f(gs))=\sigma_J$, the values chosen for
$\sigma_J$ are $3$, $4$, $5$ and $6$ as stated in Section 3.4. With increasing
$\zeta$ and $\sigma_J$ values, it is easy to see that the NTME $M^{0\nu}$ will
increase. The values of NTME for $\zeta=0.6$ and $\sigma_J=3$, 4, 5 and 6 are 1,
1.54, 2.15 and 2.83 respectively. Similarly, for $\zeta=0.65$, $0.7$ and $0.8$
they are $(1.18,1.82,2.54,3.34)$, $(1.38,2.12,2.97,3.9)$ and
$(1.78,2.74,3.82,5.03)$ respectively. With these and using $\sigma_J \sim 3-4$
and $\zeta  \sim 0.7-0.8$ will give $M^{0\nu} \sim 2-3$ in SDM while the shell
model value given in  \cite{Se82} by Horoi et al., using JUN45 interaction and
same Jastrow parameters, is 2.59. It is important to note that the shell model
results include a more detailed transition operator and other modifications. In
addition, with a different interaction Poves et al. \cite{Poves} obtained the
shell model value to  be $\sim 2.18$. As already stated in the introduction,
with other nuclear models $M^{0\nu} \sim 3-6$. Thus, it is plausible to conclude
that SDM is useful for calculating NTME for $0\nu \beta\beta$. For further
confirmation of this, in a second example $^{76}$Ge  is considered and
GERDA+MAJORANA experiments will measure the $^{76}$Ge $0\nu \beta \beta$ decay
half life in future \cite{dbd-expt2}.

For $^{76}$Ge to $^{76}$Se NDBD NTME, same shell model inputs are used as above
and similarly the parameters in the transition operator. Number of positive
parity proton-neutron  configurations is 958 for $^{76}$Ge and 2604 for
$^{76}$Se. The $\overline{\sigma}$ for $^{76}$Ge configurations is 4.4 MeV  with
a 6\% fluctuation. Similarly, for $^{76}$Se, $\overline{\sigma} = 5.51$ MeV with
a 4\% fluctuation. For the ground state determination we have \cite{nndc},  $(E_R
, J_R^\pi , N_R) = (2.02 \mbox{MeV},\, 4^+,\,37)$ for $^{76}$Ge and  $(E_R ,
J_R^\pi , N_R) = (1.79 \mbox{MeV},\, 2^+,\,33)$ for $^{76}$Se. The ground states
here are also $\sim 3 \sigma$ below the lowest configuration centroid. With all
these, the NTME are calculated and their values for $\zeta=0.65$ and
$\sigma_J=3$, 4 , 5 and 6 are 1.02, 1.56, 2.19 and 2.87 respectively. Similarly,
for $\zeta=0.7$ and $0.8$ they are $(1.29, 1.98, 2.77, 3.63)$ and $(1.96, 3.01,
4.21, 5.54)$ respectively. Shell model result from Horoi et al. \cite{Ge76},
obtained using JUN45 interaction and same Jastrow parameters, is 2.72  while it
is 2.3 from Poves et al. \cite{Poves} shell model calculations. Clearly, the
SDM values with $\zeta \sim 0.7-0.8$ and $\sigma_j \sim 4$ are close to the
shell model results.

\section{Conclusions and future outlook}

In the present paper SDM for calculating NTME for NDBD is described with all the
relevant equations. As first examples, results for $^{82}$Se and $^{76}$Ge are
presented and the SDM results are seen to be close to the shell model values.

It is clearly important that the SDM formulation given in Section 3 should be
tested. This is possible by constructing complete shell model Hamiltonian
matrix, in the configuration-$J$ basis, for the  parent and daughter nuclei
(with $J^\pi$ values fixed) and the transition  matrix generated by the action
of the transition operator on each of the  parent states taking to the daughter
states. Although this might seem complicated for realistic nuclei, a pseudo NDBD
nucleus such as  $^{24}$Mg could used for the test.

Another direction for a better SDM calculation is to evaluate all the
configuration centroids and variances with fixed-$J$ using for example, the
large scale computer codes  developed  recently by Sen'kov et al. \cite{Zel-1};
note that we need $E_c((\wm_p,\wm_n),J=0)$  and $\sigma((\wm_p,\wm_n),J=0)$.
However,  the methods used by Sen'kov et al. need to be extended to derive a
formula (or  a viable method for computing) $\l|\lan (\wm_p,\wm_n)_f J_f=0  \mid
\co \mid (\wm_p,\wm_n)_i J_i=0\ran\r|^2$. With these, it is possible in the near
future to apply the theory described in Section 3 without using Eq.
(\ref{eq.strn-app8}) for the calculation of NTME. 

Most important is to improve SDM theory with a better  treatment of $\zeta$
including its definition with configuration partitioning, although $\zeta$ via 
Eq. (\ref{eq.strn-app5}) and its extensions could be made tractable. In future,
this need to be addressed. 

It is useful to add that Eq. (\ref{eq.strn-app6}) easily gives the total
transition  strength sum, the sum of the strengths from all states of the parent
nucleus to all the states of the daughter nucleus and this depends only on the
sp space  considered. For the $^{82}$Se the total strength sum is 31239 and for
$^{76}$Ge it is 54178. Thus, $M^{0\nu}(0^+)$ is a very small fraction of the
total strength generated by the NDBD transition operator. Starting from Eq.
(\ref{eq.strn-app3}), it is  possible to obtain the total strength (NEWSR) 
originating from the ground state of the parent nucleus and also the linear and
quadratic energy weighted strength sums. These may prove to be useful in putting
constraints on the nuclear models being used for NDBD studies. This will be
addressed in future.

Finally, using SDM \cite{KH,KOKA} it is possible to study orbit occupancies and
GT distributions in various NDBD nuclei. These results can be compared with
available experimental data and will provide tests for the goodness of SDM for
NDBD. Results of these studies as well as the $M^{0\nu}$ results for heavier
$^{124}$Sn,  $^{130}$Te and $^{136}$Xe nuclei are in the process.

\ed